\documentclass[reprint,amsmath,amssymb,aps,hidelinks,pra]{revtex4-2}

\usepackage[table]{xcolor} 

\usepackage{graphicx}
\usepackage{float}
\usepackage{mathrsfs}
\usepackage{dcolumn}
\usepackage{bm}
\usepackage{hyperref}
\usepackage{orcidlink}
\hypersetup{colorlinks=true,linkcolor=blue,urlcolor=blue,citecolor=blue}
\usepackage{siunitx}

\definecolor{visbg}{HTML}{06D6A0}
\definecolor{thbg}{HTML}{26547C}  
\definecolor{hpbg}{HTML}{FF6B6B}  
\definecolor{mbbg}{HTML}{A267AC}  

\newcommand*{\figref}[2][]{%
  \hyperref[{fig:#2}]{%
    \ref*{fig:#2}%
    \ifx\\#1\\%
    \else
      #1%
    \fi
  }%
}

\begin{document}

\preprint{APS/123-QED}

\title{Revealing stimulus-dependent dynamics through statistical complexity}

\author{Edson V. de Paula\textsuperscript{1\orcidlink{0009-0002-0969-0653}}  }
\author{Rafael M. Jungmann\textsuperscript{2, 3\orcidlink{0000-0001-8952-8888}}  }
\author{Antonio J. Fontenele\textsuperscript{4\orcidlink{0000-0001-8952-8888}}  }
\author{Leandro A. A. Aguiar\textsuperscript{2, 3\orcidlink{0000-0001-8952-8888}}}
\author{Pedro V. Carelli \textsuperscript{1\orcidlink{0000-0002-5666-9606}}} 
\author{Fernanda S. Matias \textsuperscript{5\orcidlink{0000-0002-5629-6416}}}
\author{Mauro Copelli \textsuperscript{1\orcidlink{0000-0001-7441-2858}}} \email{mauro.copelli@ufpe.br}
\author{Nivaldo A. P. de Vasconcelos\textsuperscript{4,\orcidlink{0000-0002-0472-8205}}} \email{nivaldo.vasconcelos@ufpe.br}

\affiliation{\textsuperscript{1}Departamento de Física, Universidade Federal de Pernambuco, 50670-901. Recife-PE, Brazil}
\affiliation{\textsuperscript{2}Life and Health Sciences Research Institute (ICVS), School of Medicine, University of Minho, Braga, 4710-057, Portugal}
\affiliation{\textsuperscript{3}ICVS/3B’s - PT Government Associate Laboratory, Braga/Guimarães, Portugal}
\affiliation{\textsuperscript{4}UA Integrative Systems Neuroscience Group, Department of Physics, University of Arkansas, Fayetteville, AR 72701, USA}
\affiliation{\textsuperscript{5}Instituto de Física, Universidade Federal de Alagoas. 57072-970. Maceió-AL, Brazil}
\affiliation{\textsuperscript{6}Departamento de Engenharia Biomédica, Universidade Federal de Pernambuco. 50670-901. Recife-PE, Brazil}
\date{\today}

\begin{abstract}

Advances in large-scale neural recordings have expanded our ability to describe the activity of distributed brain circuits. However, understanding how neural population dynamics differ across regions and behavioral contexts remains challenging. Here, we surveyed neuronal population dynamics across multiple mouse brain areas (visual cortex, hippocampus, thalamus, and midbrain) using spike data from local ensembles. Two complementary measures were used to characterize these dynamics: the coefficient of variation (CV), a classical indicator of spike-time variability, and statistical complexity, an information-theoretic quantifier of organizational structure. To probe stimulus-dependent activity, we segmented and concatenated recordings from behavioral experiments into distinct time series corresponding to natural image presentations, blank screens during visual task, and spontaneous activity. While the CV failed to discriminate between these conditions, statistical complexity revealed clear, stimulus-specific motifs in population activity. These results indicate that information-theoretic measures can uncover structured, stimulus-dependent patterns in neural population dynamics that remain unobserved in traditional variability metrics.

\end{abstract}

\maketitle

\section{Introduction} \label{sec:Introduction}
The mammalian brain represents one of nature's most complex dynamical systems, as evidenced by the electrical activity generated by neuronal signaling \cite{Adrian1926}. Individual neurons, through their interactions within circuits give rise to rich collective behaviors, such as transitions between asynchronous, synchronous, and chaotic regimes \cite{Van_Vreeswijk1996, Van_Vreeswijk1998, Harris2011}. The complexity of cortical dynamics becomes manifest when we consider the collective properties that emerge from large neuronal ensembles, where fundamental principles governing individual circuit elements give rise to qualitatively distinct network-level phenomena~\cite{Sompolinsky2014}. 

Recent experimental advances provide direct access to the high-dimensional phase space of brain-wide activity patterns, allowing us to probe the collective properties of these neural population dynamics~\cite{Stevenson2011, hong2019, urai2022}. This new window into the brain has revealed the intrinsic relationship between spontaneous behavioral variability and the underlying multidimensional structure of neural population states~\cite{Stringer2019}, elucidated the distributed nature of decision-making and attentional modulation across cortical and subcortical networks~\cite{Steinmetz2019}, and enabled the characterization of metastable brain states and their transition dynamics~\cite{Kringelbach2020}. Ultimately, these population-level measurements help establish a theoretical framework for understanding neural computation through the lens of dynamical systems theory and statistical physics.

The Allen Institute's Visual Coding—Neuropixels dataset provides an ideal testbed for evaluating neuronal population dynamics across distributed brain networks~\cite{Durand2023, Siegle2021}. This dataset captures high-resolution neural recordings from the mouse brain using Neuropixels 1.0 probes, simultaneously recording spiking activity across multiple interconnected regions, such as the visual cortex, hippocampus, thalamus, and midbrain. These extensive recordings, taken from both wild-type and transgenic mice during standardized visual stimulation, encompass spiking data from over 300,000 neurons across distributed functional systems.

While local neuronal population spiking variability already serves as a robust metric for investigating the dynamics of mammalian cortical circuits~\cite{Renart2010, Harris2011}, the simultaneous recording of activity across multiple brain regions allows us to extend this variability-based approach to address large-scale, systems-level questions about how different brain areas coordinate their dynamics. In the present work, we apply the complexity-entropy (C-H) plane framework~\cite{Rosso2007} to quantify how ordinal information processing strategies vary systematically across the mouse brain under different behavioral and stimulus conditions. This analysis will provide new insights into the computational principles underlying distributed neural computation, specifically those based on the complexity of ordinal patterns observed in brain activity~\cite{Lucas2021, Lotfi2021, Jungmann2023}.

\section{Methods} \label{sec:Methods}

\subsection{Data Analysis}

The dataset used in this study was sourced from the Allen Institute \cite{allen2024visualbehavior}, comprising simultaneous recordings from multiple brain regions during various experimental and behavioral conditions using Neuropixels 1.0 probes (Fig. \ref{fig:cv-intro}(a)). Our analysis focuses on neural dynamics during two principal phases: active behavior and spontaneous activity.

The active behavior phase constitutes the first hour of the experiment. During this period, trained animals performed a change recognition task, receiving water rewards for correctly identifying an image change. Each session utilized a set of eight natural images presented in a repeating sequence: a 250 ms image presentation followed by a 500 ms gray screen (referred to as ``blanks"), see Fig.\ref{fig:cv-intro}(a).

We performed specific temporal manipulations on the continuous recordings to isolate and investigate neural dynamics under four distinct stimulus conditions. 

Active Behavior: The approximately 3600 seconds of continuous recording were divided into 36 non-overlapping 100-second windows (sessions shorter than 3600 s yielded fewer windows). This segment captures neural dynamics during the mixture of image and blank presentations characteristic of the active task.

Natural Images: To isolate the response to visual stimuli, we concatenated the neural spikes recorded exclusively during the presentation windows of each of the eight natural images. This yielded eight continuous time series, each approximately 140 seconds long, from which the central 100 seconds were extracted for analysis.

Blanks: To isolate activity during stimulus transitions, we concatenated spikes from the 500 ms gray screen windows that followed each image presentation. This yielded eight time series, each approximately 240 seconds long, from which the central 100 seconds were again extracted.

Spontaneous Activity: Following the active behavior phase, animals were presented with synthetic images without reward. Within this interval, we analyzed neural activity during two distinct 100-second windows sampled from a 5-minute gray screen period, representing a lack of external stimulation.

Our analysis compares neuronal activity across the four defined stimulus conditions: natural images, active behavior, blank screens, and spontaneous activity. To ensure statistical validity, we only included populations with at least 30 recorded neurons sampled across a minimum of 8 recording sessions.
All primary analyses were performed on 100-second time windows of local population activity, serving as the standardized temporal scale for all calculated metrics.

\begin{table}[ht]
\centering
\begin{tabular}{p{3.5cm}p{3cm}p{2.2cm}}
\hline
\textbf{Subarea} & \textbf{Acronym} & \textbf{Neurons}\\
\hline
\rowcolor{visbg!20}primary visual cortex & VISp & 10010\\
\rowcolor{visbg!20}lateral area & VISl & 9255\\
\rowcolor{visbg!20}anterolateral area & VISal & 8572\\
\rowcolor{visbg!20}posteromedial area & VISpm & 10250\\
\rowcolor{visbg!20}anteromedial area & VISam & 8566\\
\rowcolor{visbg!20}rostrolateral area & VISrl & 7332\\
\hline
\rowcolor{visbg!20}\textbf{visual cortex} &  & \textbf{53985}\\
\hline
\rowcolor{thbg!20}dorsal part of lateral geniculate complex & LGd & 2169\\
\rowcolor{thbg!20}lateral posterior nucleus of the thalamus & LP & 2795\\
\rowcolor{thbg!20}posterior intralaminar thalamic nucleus & PIL & 868\\
\rowcolor{thbg!20}posterior limiting nucleus of the thalamus & POL & 1298\\
\rowcolor{thbg!20}medial geniculate complex, dorsal part & MGd & 2497\\
\rowcolor{thbg!20}medial geniculate complex, ventral part & MGv & 4492\\
\rowcolor{thbg!20}medial geniculate complex, medial part & MGm & 1928\\
\rowcolor{thbg!20}suprageniculate nucleus & SGN & 3033\\
\rowcolor{thbg!20}posterior triangular thalamic nucleus & PoT & 1461\\
\hline
\rowcolor{thbg!20}\textbf{thalamic regions} &  & \textbf{20541}\\
\hline
\rowcolor{hpbg!20}dentate gyrus & DG & 7486\\
\rowcolor{hpbg!20}CA3 field & CA3 & 3729\\
\rowcolor{hpbg!20}CA1 field  & CA1 & 22399\\
\rowcolor{hpbg!20}prosubiculum & ProS & 3067\\
\rowcolor{hpbg!20}subiculum & SUB & 4348\\
\rowcolor{hpbg!20}postsubiculum & POST & 1367\\
\hline
\rowcolor{hpbg!20}\textbf{hippocampus} &  & \textbf{42396}\\
\hline
\rowcolor{mbbg!20}nucleus of the optic tract & NOT & 309\\
\rowcolor{mbbg!20}nucleus of the brachium of the inferior colliculus & NB & 793\\
\rowcolor{mbbg!20}midbrain reticular nucleus & MRN & 3661\\
\rowcolor{mbbg!20}superior colliculus, intermediate gray layer & SCig & 1684\\
\rowcolor{mbbg!20}anterior pretectal nucleus & APN & 10716\\
\hline
\rowcolor{mbbg!20}\textbf{midbrain areas} &  & \textbf{17163}\\
\hline
\end{tabular}
\caption{Local populations abbreviations and number of neurons analysed.}
\label{tab:visual_areas}
\end{table}

\subsection{Population Firing Rate and Variability Analysis}

We quantified the population instantaneous firing rate $r(t)$ by convolving spike trains across all recorded units with a sliding temporal window. For each time point $t$, the population firing rate was computed as:
\begin{equation}
r(t) = \frac{1}{N\Delta T} \int_{t}^{t+\Delta T} \rho(\tau) \, d\tau,
\label{eq:firing_rate}
\end{equation}
where $N$ denotes the total number of recorded neurons, $\rho(t) = \sum_{j=1}^{N} \sum_{k} \delta(t - t_j^k)$ represents the population spike train (with $t_j^k$ indicating the $k$-th spike time of neuron $j$), and $\Delta T = 50$~ms is the integration window (unless otherwise stated).

To quantify temporal variability in population activity, we calculated the coefficient of variation (CV) of the firing rate time series. Each recording epoch was divided into non-overlapping 10-second windows, and within each window $i$, the CV was computed as:
\begin{equation}
\text{CV}_i = \frac{\sigma_i}{\mu_i},
\label{eq:cv}
\end{equation}
where $\mu_i$ and $\sigma_i$ represent the mean and standard deviation of $r(t)$ within the $i$-th window, respectively. For each 100-second data epoch, the final CV was obtained by averaging the ten $\text{CV}_i$ values across constituent windows.

\subsection{Bandt and Pompe Symbolization Technique}

We characterized the temporal dynamics of population firing rates using the symbolic ordinal pattern method developed by Bandt and Pompe~\cite{Bandt2002}. This approach captures the temporal ordering structure of time series data while remaining invariant to monotonic transformations and robust to observational noise.

For a univariate time series $\{X(t): t = 1, \ldots, T\}$ of length $T$, we specified an embedding dimension $D \geq 2$ and time delay $\tau \geq 1$ (with $D, \tau \in \mathbb{N}$). At each time index $t$, we constructed a delay-embedded vector:
\begin{equation}
\mathbf{s}(t) = \left(x_t, x_{t+\tau}, x_{t+2\tau}, \ldots, x_{t+(D-1)\tau} \right).
\label{eq:delay_vector}
\end{equation}
Each vector $\mathbf{s}(t)$ was mapped to a unique ordinal pattern $\pi_j = (r_0, r_1, \ldots, r_{D-1})$, where $(r_0, r_1, \ldots, r_{D-1})$ represents a permutation of $(0, 1, \ldots, D-1)$ encoding the rank order of elements in $\mathbf{s}(t)$. Specifically, the pattern $\pi$ satisfies:
\begin{equation}
x_{t+r_0 \tau} \leq x_{t+r_1 \tau} \leq \cdots \leq x_{t+r_{D-1} \tau}.
\label{eq:ordinal_pattern}
\end{equation}
In cases of equal values (ties), we assigned higher ranks to later-occurring elements to ensure deterministic pattern assignment.

The frequency of occurrence for each ordinal pattern was computed by sliding the embedding window across the entire time series, yielding $T - (D-1)\tau$ pattern observations. These frequencies were normalized to obtain a probability distribution $P \equiv \lbrace p_j; j = 1, 2, ..., D! \rbrace$ over the $D!$ possible ordinal patterns. To ensure reliable statistical estimation, the condition $T \gg D!$ must be satisfied. Following established guidelines~\cite{Bandt2002}, we set $D = 6$ and $\tau = 1$ throughout all analyses, resulting in 720 possible ordinal patterns. Given our 100-second analysis windows with 50~ms temporal bins, each time series contained $T = 2000$ data points, providing sufficiently well-distributed ordinal-pattern estimates.

\subsection{Information-Theoretic Quantifiers}

To characterize the information-theoretic properties of neural population dynamics, we computed two complementary measures from the ordinal pattern distributions: Shannon entropy and statistical complexity.

Shannon entropy~\cite{Shannon1949} quantifies the degree of unpredictability in the ordinal pattern distribution and is defined as:
\begin{equation}
S[P] = -\sum_{j=1}^{D!} p_j \ln(p_j),
\label{eq:shannon_entropy}
\end{equation}
where $p_j$ denotes the probability of the $j$-th ordinal pattern. To enable comparison across different embedding dimensions and experimental conditions, we normalized the entropy by its theoretical maximum, which occurs for a uniform distribution $P_e = \{p_j = 1/D!,\ \forall j\}$:
\begin{equation}
H[P] = \frac{S[P]}{S[P_e]} = \frac{S[P]}{\ln(D!)},
\label{eq:normalized_entropy}
\end{equation}
where $0 \leq H[P] \leq 1$. Values near zero ($H \approx 0$) indicate highly deterministic or periodic dynamics with few dominant patterns, while values approaching unity ($H \approx 1$) reflect stochastic or unpredictable dynamics with a nearly uniform pattern distribution.

While entropy measures randomness, it does not distinguish between different types of structural organization. Statistical complexity~\cite{Lamberti2004} addresses this limitation by quantifying the degree of correlated structure present in the time series. It is defined as the product of two factors:
\begin{equation}
C[P] = Q[P, P_e] \cdot H[P],
\label{eq:complexity}
\end{equation}
where $Q[P, P_e]$ represents the disequilibrium between the empirical pattern distribution $P$ and the uniform reference distribution $P_e$. This disequilibrium is computed as:
\begin{equation}
Q[P, P_e] = Q_0 \cdot \mathscr{D}_{JS}[P, P_e],
\label{eq:disequilibrium}
\end{equation}
where $Q_0$ is a normalization constant ensuring $0 \leq Q \leq 1$, and $\mathscr{D}_{JS}$ denotes the Jensen--Shannon divergence as
%
\begin{equation}
\mathscr{D}_{JS}[P_1, P_2] = S\left[\frac{P_1 + P_2}{2}\right] \\
- \frac{S[P_1]}{2} - \frac{S[P_2]}{2}.
\label{eq:js_divergence}
\end{equation}
%


Statistical complexity achieves its minimum value ($C = 0$) for both completely ordered (single pattern) and completely random (uniform distribution) dynamics, as both lack meaningful structure~\cite{Martin2006, Lopez1995}. Intermediate values of $C$ indicate the presence of correlated temporal patterns, with maximum complexity occurring for systems exhibiting a balance between order and disorder~\cite{Martin2006}.

The joint representation of normalized permutation entropy $H[P]$ and statistical complexity $C[P]$ in a two-dimensional plane forms the complexity-entropy (C-H) causality plane~\cite{Rosso2007}. This framework enables the classification and visualization of different dynamical regimes, ranging from deterministic and periodic dynamics (low $H$, low $C$) to stochastic processes (high $H$, low $C$) and complex correlated structures (intermediate $H$, high $C$) Fig. ~\ref{fig:cplx-intro}(a).

\subsection{Surrogate data}

To verify the robustness of our results, we generated surrogate data for each analyzed time window by independently shuffling the inter-spike intervals (ISIs) of each neuron. This procedure preserved each unit's firing rate distribution while destroying temporal correlations and coordinated activity across the population.

\begin{figure*}[p]
    \centering
    \includegraphics[width=\textwidth]{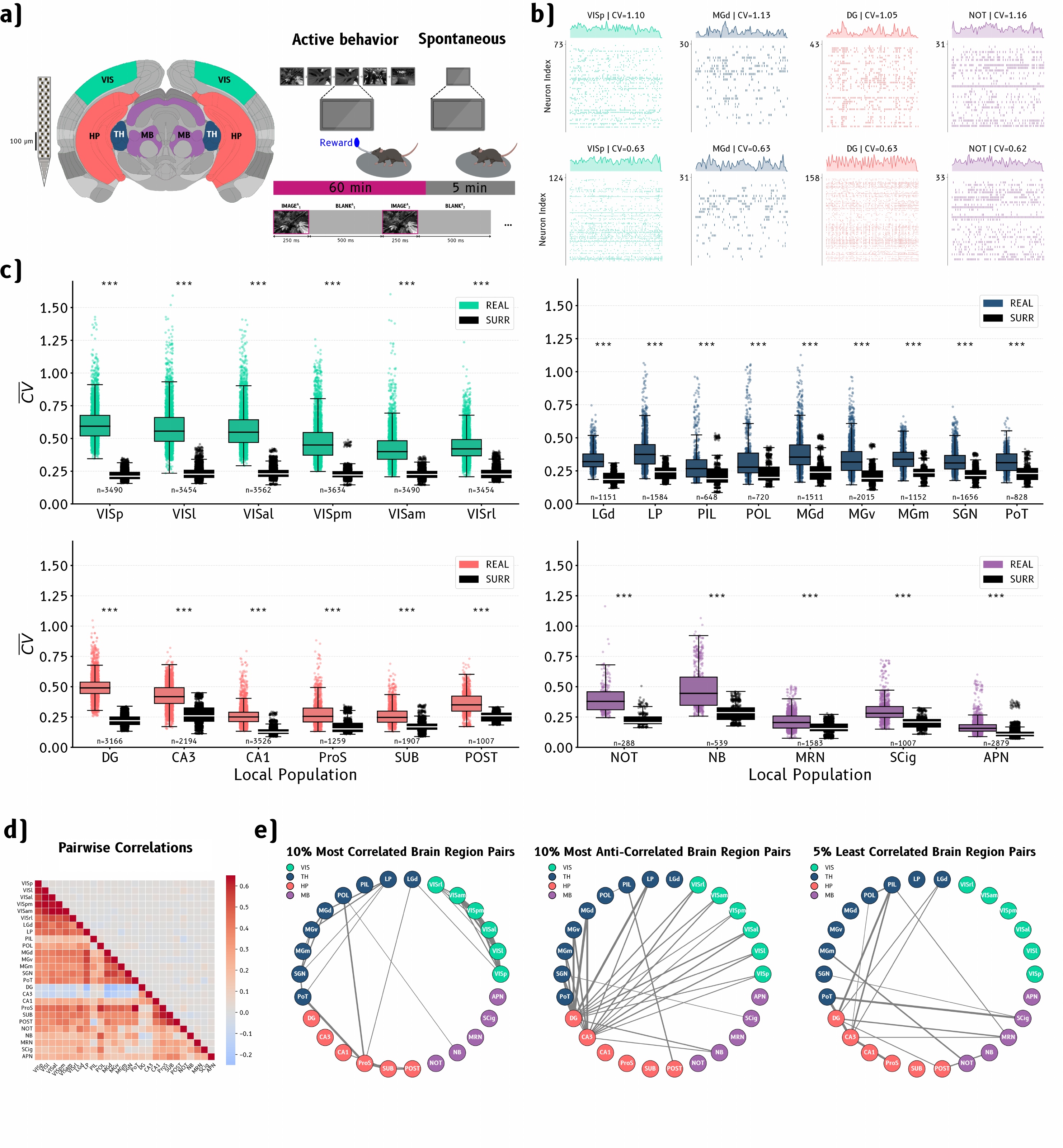}
    \caption{
\textbf{Coefficient of variation reveals region-specific population dynamics and inter-regional coordination during active behavior.}
\textbf{(a)} Experimental setup. Left: Neuropixels 1.0 probe. Middle: Coronal section showing visual cortex (VIS, green), thalamus (TH, blue), hippocampus (HP, coral), and midbrain (MB, purple). Right: Behavioral task during active behavior epoch. Animals viewed natural images (250~ms each) followed by 500~ms blank screens, reporting detected changes via licking for water reward.
\textbf{(b)} Representative raster plots showing different population variability levels across brain regions during 4-second windows, with population firing rates above each raster. 
\textbf{(c)} CV distributions across brain regions during active behavior ($n = 102$ sessions). Colored boxplots: empirical data; black boxplots: temporally shuffled surrogates. Visual cortex subregions showed highest CV values (median: 0.40--0.59), 2--3$\times$ higher than surrogates. Thalamic regions exhibited moderate values (median: 0.27--0.37), 1.3--1.6$\times$ above surrogates. Hippocampal subregions displayed high CV (median: 0.25--0.49), 1.5--2.2$\times$ higher than surrogates, with dentate gyrus showing maximum values. Midbrain areas had lowest CV (median: 0.16--0.45), 1.3--1.9$\times$ higher than surrogates. Mann--Whitney tests confirmed significant differences for all regions ($p < 0.001$).
\textbf{(d)} Pairwise Pearson correlation matrix of CV time series using 10-second non-overlapping windows. Lower triangle: empirical correlations; upper triangle: surrogate correlations.
\textbf{(e)} Network graphs of inter-regional CV relationships. Left: Top 10\% positive correlations (0.50--0.67), showing functional clustering within visual cortex, hippocampus, and thalamus. Middle: Top 10\% anticorrelations ($-0.27$ to $-0.002$), highlighting dentate gyrus and CA3 as anticorrelation hubs. Right: Near-zero correlations ($-0.04$ to $+0.03$), showing midbrain independence from cortical and hippocampal networks.
    }
    \label{fig:cv-intro}
\end{figure*}

\section{Results} \label{sec:Results}
\subsection{Spiking variability during active behavior}

Natural images reveal coding principles and functional organization in visual circuits that simplified stimuli cannot capture, while driving complex population dynamics across distributed brain networks~\cite{van-Hateren1998, Simoncelli2001, Stringer2019}. However, the temporal structure and coordination of population responses to natural images across distributed brain regions remain poorly understood. In particular, while population dynamics and their coordination have been characterized in cortical areas~\cite{Vasconcelos2017}, such analyses are lacking for other regions involved in visual processing. In cortical circuits, the spiking variability of local neuronal populations has been used as a proxy for their collective state. To address this gap, we quantified the spiking variability of local neuronal populations across the four major brain regions described above (visual cortex, thalamus, hippocampus, and midbrain) by computing the coefficient of variation (CV) of their population firing rates~\cite{Renart2010,Harris2011} during natural image presentation, as shown in Fig.\ref{fig:cv-intro}(a). Data were acquired using six Neuropixels probes inserted at different locations targeting the visual cortex.

Figure \ref{fig:cv-intro}(b) shows raster plots of local neuronal populations across regions at different CV levels. Different levels of CV in primary sensory areas have been associated with different degrees of synchrony in local neuronal populations~\cite{Harris2011, Renart2010}. Moreover, CV time series display strong correlations across primary sensory areas~\cite{Vasconcelos2017}. This consistency breaks down when examining neuronal populations beyond primary sensory regions.

Figure~\ref{fig:cv-intro}(c) shows the $\overline{CV}$ distributions across brain areas (all anatomical abbreviations are defined in Table~\ref{tab:visual_areas}) during active behavior (colored boxplots) compared with surrogate data distributions obtained by randomly shuffling spike times within each neuron. The analysis revealed region-specific patterns of population variability. Visual cortex subregions exhibited the highest CV values (median range: 0.40--0.59, interquartile range (IQR): 0.12--0.18), approximately 2--3$\times$ higher than their shuffled surrogates (median range: 0.21--0.24). Thalamic regions showed moderate CV values (median range: 0.27--0.37, IQR: 0.10--0.15), 1.3--1.6$\times$ higher than surrogates. Hippocampal subregions displayed high CV values (median range: 0.25--0.49, IQR: 0.08--0.13), 1.5--2.2$\times$ higher than surrogates, with the dentate gyrus showing the maximum median CV (0.49). Midbrain areas exhibited the lowest CV values (median range: 0.16--0.45, IQR: 0.05--0.23), 1.3--1.9$\times$ higher than surrogates. Mann--Whitney tests confirmed significant differences between empirical and surrogate distributions for all brain regions ($p < 0.001$), indicating that the observed CV patterns reflect genuine temporal structure in neural population activity rather than arising from independent neuronal firing. 

Beyond regional differences, we  examined inter-regional relationships by constructing a CV correlation matrix between brain areas during active behavior (Fig.~\ref{fig:cv-intro}(d), lower triangle). Correlations were computed using CV values from non-overlapping 10-second windows to ensure statistical independence of observations. To further elucidate these connectivity patterns, we generated the corresponding network graphs identifying the strongest CV relationships (Fig.~\ref{fig:cv-intro}(e)). The left panel highlights the 10\% most positively correlated region pairs, revealing preferential connections within functional modules: visual areas clustered together, as did subicular regions and thalamic populations. Notably, connections between functionally distinct brain systems were largely absent. The middle panel presents the 10\% most anticorrelated pairs, emphasizing DG and CA3 as a hub of negative CV relationships with other regions. The right panel displays near-zero correlations, showing that midbrain areas exhibited the weakest CV correlations across all studied regions, suggesting relative functional independence from cortical and hippocampal dynamics. For comparison, the upper triangle displays correlations computed from surrogate data, demonstrating that shuffling spike times abolished inter-regional CV relationships. 
 


\subsection{Cortical but not subcortical dynamics are stimulus-dependent}

\begin{figure*}[p]
    \centering
    \includegraphics[width=\textwidth]{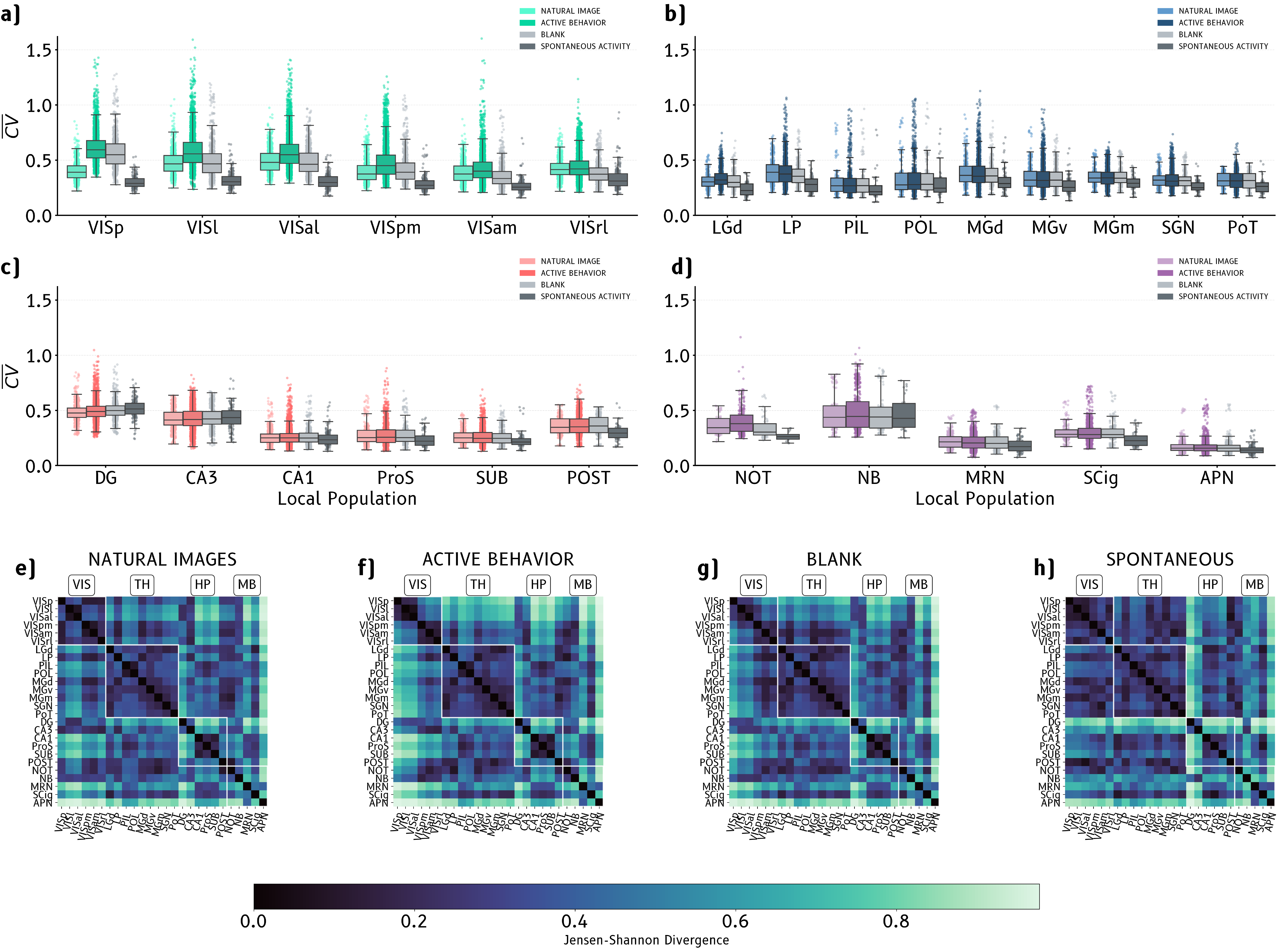} 
    
    \caption{
    \textbf{Coefficient of variation exhibits stimulus-dependent modulation in cortical but not subcortical regions.}
    \textbf{(a)} Distribution of CV values across experimental conditions for all brain regions ($n = 102$ sessions, total 77,216 analysis windows). Boxplots show CV distributions during four behavioral states: natural images (light-colored boxes, image only presentations, $n = 11,321$ windows total), active behavior (dark-colored boxes, $n = 51,704$ windows), blank screens (light-gray boxes, inter-image gray screen presentation, $n = 11,321$ windows), and spontaneous activity (dark-gray boxes, gray screen as visual input, $n = 2,870$ windows). Visual cortex subregions exhibited distinct CV distributions across conditions, with active behavior and blank screens showing elevated median CV values (VISp: 0.59 and 0.55; VISl: 0.56 and 0.46; VISal: 0.55 and 0.46, respectively) compared to spontaneous activity (VISp: 0.29; VISl: 0.30; VISal: 0.30) and natural images (VISp: 0.39; VISl: 0.46; VISal: 0.48). Thalamic regions showed similar patterns with reduced effect sizes (e.g., LP median CV: 0.37 in active behavior vs. 0.28 in spontaneous activity; LGd: 0.32 vs. 0.23). In contrast, hippocampal subregions displayed relatively stable median CV values across visually driven conditions (DG: 0.48--0.50; CA3: 0.41--0.42; CA1: 0.25) but showed modest increases during spontaneous activity for DG (0.51) and CA3 (0.43). Midbrain areas exhibited the most invariant distributions across all conditions (e.g., APN median CV: 0.16 across all states; MRN: 0.20--0.21), suggesting stimulus-independent dynamics.
    \textbf{(b)} Jensen--Shannon (JS) divergence matrices quantifying the dissimilarity between CV distributions across pairs of brain region for each experimental conditions. Color scale ranges from 0.0 (black) to 0.98 (light green). Visual cortex and thalamic regions showed large JS divergences between spontaneous activity and visually driven states, with values exceeding 0.40 for most visual areas and 0.30 for thalamic regions. Hippocampal regions, particularly DG and CA3, exhibited moderate divergences (0.15--0.35) across condition pairs, with the largest values occurring between spontaneous and active behavior conditions. Among the three visually driven conditions (natural images, active behavior, blank screens), JS divergences were consistently low across cortical regions (typically $< 0.20$), indicating comparable population variability dynamics despite differences in task engagement and visual content. Midbrain areas displayed the smallest divergences overall ( JS $< 0.15$ for most condition pairs), confirming their relative insensitivity to stimulus conditions.}
    \label{fig:cv-stimuli}
\end{figure*}

Having established that different brain regions exhibit distinct baseline levels of population variability, an important question emerges: how does this variability respond to changes in stimulus conditions and behavioral state? Specifically, it reveals whether population variability patterns reflect intrinsic properties of neural circuits~\cite{Fox2005, Fiser2004} or are instead flexibly modulated by sensory input and behavioral context~\cite{Stitt2019, Buonomano2009}?. An influential perspective proposes that the brain operates primarily through intrinsic dynamics, with sensory information modulating rather than determining the system's operation~\cite{Fox2005}. However, the extent to which this principle holds across different regions of the visual processing hierarchy remains unclear. Primary sensory regions, which directly encode stimulus features, might exhibit strong context-dependent modulation of their population variability. Conversely, regions supporting more abstract or internally generated computations, such as memory consolidation in the hippocampus or arousal regulation in the midbrain---might maintain relatively stable dynamics regardless of visual input~\cite{Grossberg2009}. To test these predictions, we compared CV distributions across four experimental conditions: natural image presentation during active behavior, blank screens during active behavior, natural images alone, and spontaneous activity without visual stimulation or task engagement (Fig.~\ref{fig:cv-stimuli}). This approach allowed us to disentangle the contributions of visual input, behavioral engagement, and intrinsic dynamics to population variability across the visual processing hierarchy.

Figure~\ref{fig:cv-stimuli}(a) displays the distributions of CV values across different behavioral states for all brain regions. Local neuronal populations in visual cortex exhibited strong stimulus-dependent modulation, with active behavior and blank screen conditions showing elevated median CV values (VISp: 0.59 and 0.55; VISl: 0.56 and 0.46; VISal: 0.55 and 0.46) compared to spontaneous activity (VISp: 0.29; VISl: 0.30; VISal: 0.30), indicating transitions between low-synchrony and high-synchrony population states. Natural image presentation alone produced intermediate CV values (VISp: 0.39; VISl: 0.46; VISal: 0.48). Thalamic local neuronal populations showed qualitatively similar patterns but with reduced effect sizes (LP: 0.37 in active behavior vs. 0.28 in spontaneous activity; LGd: 0.32 vs. 0.23), suggesting attenuated but present stimulus modulation in these subcortical sensory relays.

In contrast, hippocampal local neuronal populations displayed markedly different behavior. DG and CA3 maintained relatively stable median CV values across all visually driven conditions (DG: 0.48--0.50; CA3: 0.41--0.42), with only modest increases during spontaneous activity (DG: 0.51; CA3: 0.43). Midbrain local neuronal populations exhibited the most invariant dynamics, with essentially constant CV values across all experimental conditions (APN: 0.16; MRN: 0.20--0.21), indicating complete independence from the different stimulus conditions.

To systematically quantify these differences, we computed the Jensen-Shannon (JS) divergence between CV distributions for each pair of brain regions across different stimulus conditions [Fig.~\ref{fig:cv-stimuli}(b)]. Low JS divergence values indicate similar dynamical ranges between regions, whereas high values reflect distinct dynamical representations. To compare stimulus-dependent changes within each major brain area (VIS, TH, HP, and MB), we summed the JS divergence values along the main diagonal of each submatrix, representing within-area comparisons across all constituent subregions, and calculated the percentage difference between stimulus conditions.

Visual cortex showed similar JS divergence values during active behavior and blank screens ($\sim$1\% difference in summed divergences), while these conditions separated substantially from image presentation and spontaneous activity by $\sim$30\%, confirming substantial state-dependent reorganization of population dynamics. Within hippocampal regions, DG and CA3 exhibited high JS divergences across all four conditions, with the largest values occurring during spontaneous activity (median DG: 0.84; CA3: 0.84) and blank screens (median DG: 0.63; CA3: 0.63). The highest percentage difference in hippocampus occurred between active behavior and spontaneous activity ($\sim$15\%). Notably, midbrain areas displayed the smallest percentage differences overall across most condition pairs (maximum 7\%), confirming their relative insensitivity to stimulus conditions. Among the three visually driven conditions (natural images, active behavior, blank screens), JS divergences remained consistently low across cortical regions (typically $<$4\%), indicating that population variability dynamics remain comparable across behaviorally relevant states despite differences in visual content and task engagement. These results reveal a hierarchical organization where cortical dynamics are flexibly modulated by sensory input~\cite{Gautam2015,Akella2025}, while subcortical regions---particularly midbrain---maintain intrinsic dynamics largely independent of stimulus conditions.

\subsection{Statistical complexity reveals distinct dynamical regimes across brain regions}

\begin{figure*}[!t]
    \centering
    \includegraphics[width=\textwidth]{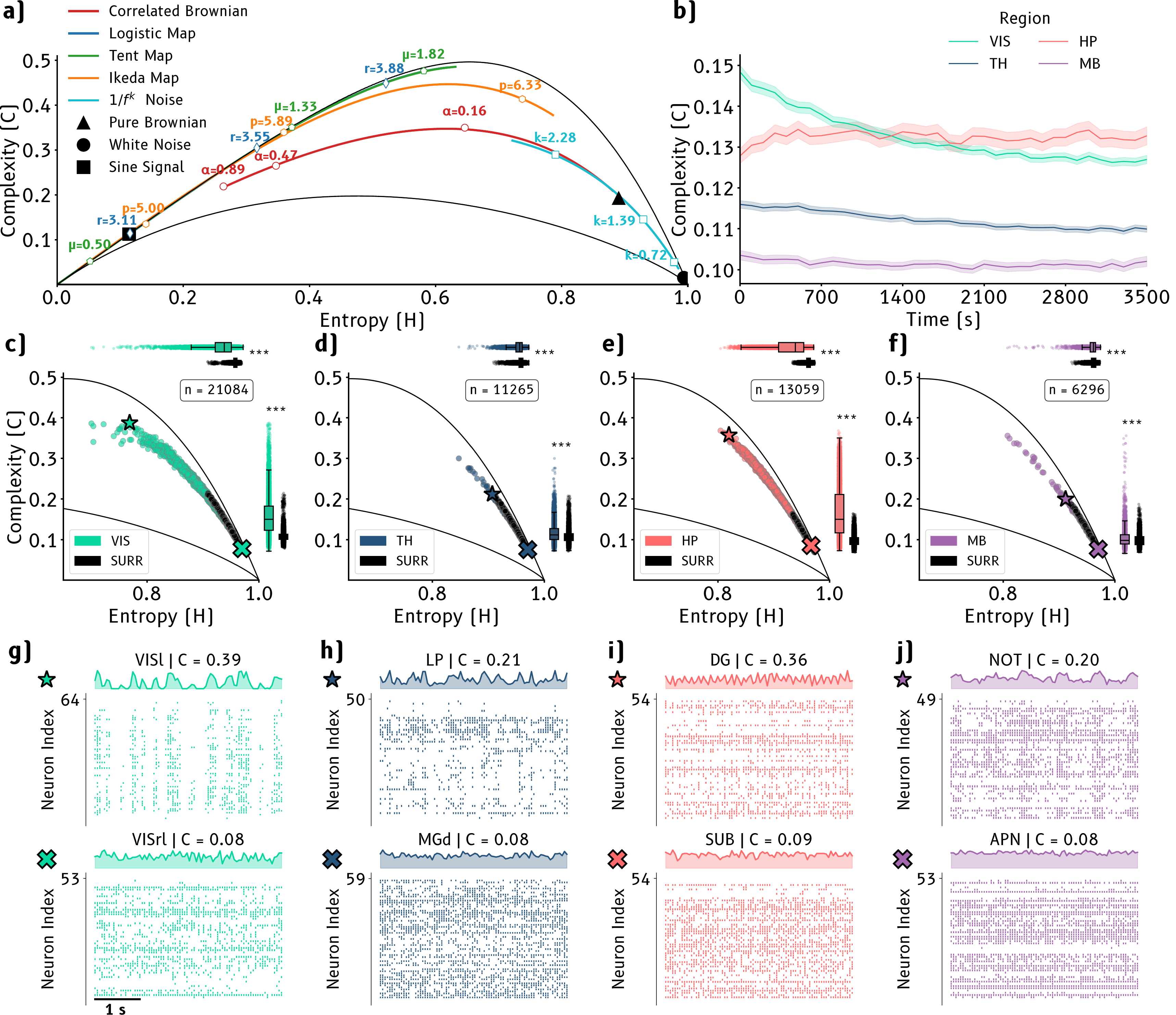} 
    \caption{
    Complexity-entropy analysis during natural image stimuli.
    \textbf{(a)} Localization of various dynamical systems and stochastic processes in the C-H plane; black boundary curves represent theoretical minimum $C_{\text{min}}$ and maximum $C_{\text{max}}$ complexity limits; different symbols denote distinct system families: pure Brownian motion (black triangle), white noise (black circle), sine signal (black square), exponentialy correlated brownian motion (orange circles with fitted curve), logistic map (pink diamonds), tent map (green pentagons), Ikeda maps (purple circles), and $1/f$ noise (green squares), with corresponding parameter values shown; $D = 6$ is used.
    \textbf{(b)} Temporal evolution of statistical complexity across different brain regions; visual cortex (VIS, green, n\textsubscript{pop} = 586), thalamus (TH, blue, n\textsubscript{pop} = 313), hippocampus (HP, coral, n\textsubscript{pop} = 363), and midbrain (MB, purple, n\textsubscript{pop} = 175); shaded areas represent standard error.
    \textbf{(c-f)} Complexity-entropy plane analysis for individual brain regions; data points represent complexity and entropy values (100 s long windows/session, bin=50 ms), bottom panels show 4-second raster plots  of neuronal populations nearby the extremes of the C-H curve;  (c) visual cortex (VIS, green, n\textsubscript{points} = 21084), (d) thalamus (TH, blue, n\textsubscript{points} = 11265), (e) hippocampus (HP, coral, n\textsubscript{points} = 13059), and (f) midbrain (MB, purple, n\textsubscript{points} = 6296). Black curves show theoretical complexity boundaries; box plots compare original data with surrogate controls (SURR, black). Asterisks indicate statistical significance (***p $<$ 0.001) between empirical and surrogate data distributions.
    \textbf{(g-j)} Raster plots showing the spike timing and variability of neuronal activity in different brain regions under varying levels of statistical complexity.
    }
    \label{fig:cplx-intro}
\end{figure*}

While the coefficient of variation provides a scalar measure of population variability that captures the degree of synchronization in local neural activity, it does not fully characterize the temporal structure and dynamical complexity of population responses. To obtain a more comprehensive characterization of population dynamics, we employed information-theoretic measures based on the Bandt-Pompe symbolization method~\cite{Bandt2002}: Shannon entropy $H$, which quantifies the degree of randomness or predictability in the time series, and Martin-Plastino-Rosso (MPR) statistical complexity $C$~\cite{Martin2006}, which captures the balance between order and disorder. Together, these measures define a complexity-entropy (C-H) plane that can distinguish different types of dynamical regimes, from purely stochastic processes to deterministic chaos~\cite{Rosso2007}. This approach has proven particularly effective for characterizing cortical dynamics at both the spiking level~\cite{Lotfi2021} and the level of local field potentials~\cite{Jungmann2024}, as different positions in this plane correspond to qualitatively distinct types of temporal organization. We computed these quantifiers for population firing rate time series across all brain regions during natural image presentation to investigate whether different regions operate in distinct dynamical regimes and whether these regimes reflect the functional specialization observed in our CV analysis (Fig.~\ref{fig:cplx-intro}).

To establish a reference framework for interpreting neural population dynamics, we first characterized the positions of canonical dynamical systems and stochastic processes in the C-H plane [Fig.~\ref{fig:cplx-intro}(a)]. As expected from theory~\cite{Rosso2007}, purely random processes occupy the extremes of low complexity: white noise approaches maximum entropy ($H \approx 1$) with minimal complexity ($C \approx 0$), while deterministic periodic signals (sine wave) exhibit minimum entropy ($H \approx 0$) and low complexity ($C \approx 0.1$). In contrast, processes with rich temporal structure occupy intermediate positions with elevated complexity. Correlated Brownian motion with varying correlation strengths traces a characteristic arc through the plane, with exponentially correlated processes (colored noise) achieving moderate complexity ($C \approx 0.3$--$0.5$) at intermediate entropy values ($H \approx 0.3$--$0.6$). Deterministic chaotic systems---including the logistic map, tent map, and Ikeda map---cluster near maximum complexity ($C \approx 0.4$--$0.5$) at high entropy ($H \approx 0.5$--$0.8$), while $1/f^k$ noise occupies a distinct trajectory reflecting its scale-free temporal correlations. These reference trajectories provide key benchmarks for interpreting the positions of neural population activity in the C-H plane.

Examining the temporal evolution of statistical complexity across brain regions during natural image presentation revealed striking regional stability [Fig.~\ref{fig:cplx-intro}(b)]. Visual cortex (VIS) and hippocampus (HP) maintained consistently high complexity values ($C \approx 0.13$--$0.15$) throughout the hour-long recording session (n=102 sessions), with modest fluctuations around their mean levels. Thalamus (TH) exhibited intermediate complexity ($C \approx 0.11$), while midbrain (MB) showed the lowest and most stable complexity values ($C \approx 0.10$). Importantly, these regional differences in complexity remained stable across time, suggesting that they reflect intrinsic properties of each region's network dynamics rather than transient fluctuations in behavioral state or arousal. The hierarchical organization observed in complexity values---highest in cortex and hippocampus, intermediate in thalamus, lowest in midbrain---mirrors the gradient in stimulus-dependent CV modulation observed in Fig.~\ref{fig:cv-stimuli}, suggesting a fundamental relationship between dynamical complexity and functional flexibility across the visual processing hierarchy. 

To characterize the full distribution of dynamical states sampled by each region, we examined the positions of individual analysis windows in the C-H plane [Fig.~\ref{fig:cplx-intro}(c--f)]. Local neuronal population in the visual cortex occupied a broad region of the plane spanning high entropy values ($H \approx 0.8$--$1.0$) with complexity values ranging from $C \approx 0.1$ to $C \approx 0.4$ [Fig.~\ref{fig:cplx-intro}(c)]. This distribution traced a characteristic arc reminiscent of correlated stochastic processes and noisy chaotic maps suggesting that visual cortical dynamics combine elements of both stochasticity and temporal structure. Crucially, the empirical distribution differed significantly from temporally shuffled surrogate data (Mann--Whitney test, $p < 0.001$), with surrogate data clustering at lower complexity values ($C \approx 0.1$--$0.2$), confirming that the observed complexity reflects genuine temporal organization rather than random spiking patterns. Representative raster plots illustrate this diversity (Fig~\ref{fig:cplx-intro}g): periods of high complexity ($C = 0.39$, VISl) showed temporally structured activity with bursts of coordinated firing, while low complexity periods ($C = 0.08$, VISrl) exhibited more irregular, desynchronized spiking patterns.

Thalamic local neuronal populations [Fig.~\ref{fig:cplx-intro}(d)] displayed a similar but more compact distribution, occupying a narrower range of the C-H plane with slightly lower entropy ($H \approx 0.75$--$0.95$) and complexity ($C \approx 0.05$--$0.25$) compared to visual cortex. The tighter distribution suggests more constrained dynamics, consistent with the thalamus's role as a sensory relay. Representative examples from LP ($C = 0.21$) and MGd ($C = 0.08$) (Fig~\ref{fig:cplx-intro}h) demonstrate this range, with higher complexity associated with more structured temporal patterning. As in visual cortex, empirical data differed significantly from surrogates ($p < 0.001$), indicating that thalamic complexity reflects intrinsic network properties.

Hippocampal populations [Fig.~\ref{fig:cplx-intro}(e)] exhibited the broadest distribution, spanning an extensive range of complexity ($C \approx 0.05$--$0.4$) at high entropy ($H \approx 0.85$--$1.0$). This wide dispersion suggests that hippocampal circuits can flexibly transition between diverse dynamical regimes, potentially reflecting their involvement in multiple cognitive functions including spatial navigation and memory processing. High complexity states (DG: $C = 0.36$; SUB: $C = 0.09$) showed markedly different temporal structures in their raster plots, with DG exhibiting more organized sequential activation patterns. The significant difference from surrogates ($p < 0.001$) confirms that this dynamical diversity reflects genuine network computations.

Midbrain populations [Fig.~\ref{fig:cplx-intro}(f)] occupied the most restricted region of the C-H plane, clustering at high entropy ($H \approx 0.9$--$1.0$) but consistently low complexity ($C \approx 0.05$--$0.2$). This narrow distribution suggests relatively simple, near-random dynamics with limited temporal structure. Representative examples from NOT ($C = 0.20$) and APN ($C = 0.08$) showed largely irregular firing patterns with minimal temporal organization. Nevertheless, empirical data remained significantly different from surrogates ($p < 0.001$), indicating some degree of coordinated activity beyond pure randomness.

Collectively, these results reveal a hierarchical organization of dynamical complexity that parallels the functional specialization across the visual processing hierarchy. Visual cortex and hippocampus explore broad regions of the C-H plane, suggesting rich repertoires of dynamical states that may support flexible stimulus processing and memory functions. Thalamus occupies an intermediate position, while midbrain dynamics remain close to random processes, consistent with their more stereotyped, stimulus-independent responses observed in the CV analysis. The systematic differences from surrogate data across all regions confirm that these complexity signatures reflect genuine temporal coordination in population activity rather than trivial statistical artifacts.

Representative raster plots spanning the range of complexity values observed in each region [Fig.~\ref{fig:cplx-intro}(g)] illustrate how statistical complexity relates to the temporal structure of population activity. In visual cortex, high complexity states (VISl: $C = 0.39$) exhibited coordinated bursts of activity with clear temporal patterning, while low complexity states (VISrl: $C = 0.08$) showed sparse, irregular firing. Thalamic regions displayed moderate temporal organization at intermediate complexity (LP: $C = 0.21$; MGd: $C = 0.08$), with discernible differences in burst synchronization. Hippocampal populations showed pronounced contrasts, with high complexity examples (DG: $C = 0.36$) revealing sequential activation patterns and dense, structured firing, whereas lower complexity states (SUB: $C = 0.09$) exhibited more scattered activity. Midbrain areas consistently displayed irregular, weakly coordinated firing regardless of complexity level (NOT: $C = 0.20$; APN: $C = 0.08$), reflecting their restricted position in the C-H plane. However, these aggregate measures of complexity, like CV, do not fully capture the heterogeneity evident in the raster plots. Notably, subcortical populations rarely exhibit complete silence, with persistent asynchronous activity across neurons even during low complexity states, suggesting that multiple sub-dynamics coexist within local neuronal populations that are not entirely reflected in scalar measures of temporal organization.

\subsection{Stimulus-dependent modulation of dynamical complexity mirrors functional hierarchy}

\begin{figure*}[p]
    \centering
    \includegraphics[width=\textwidth]{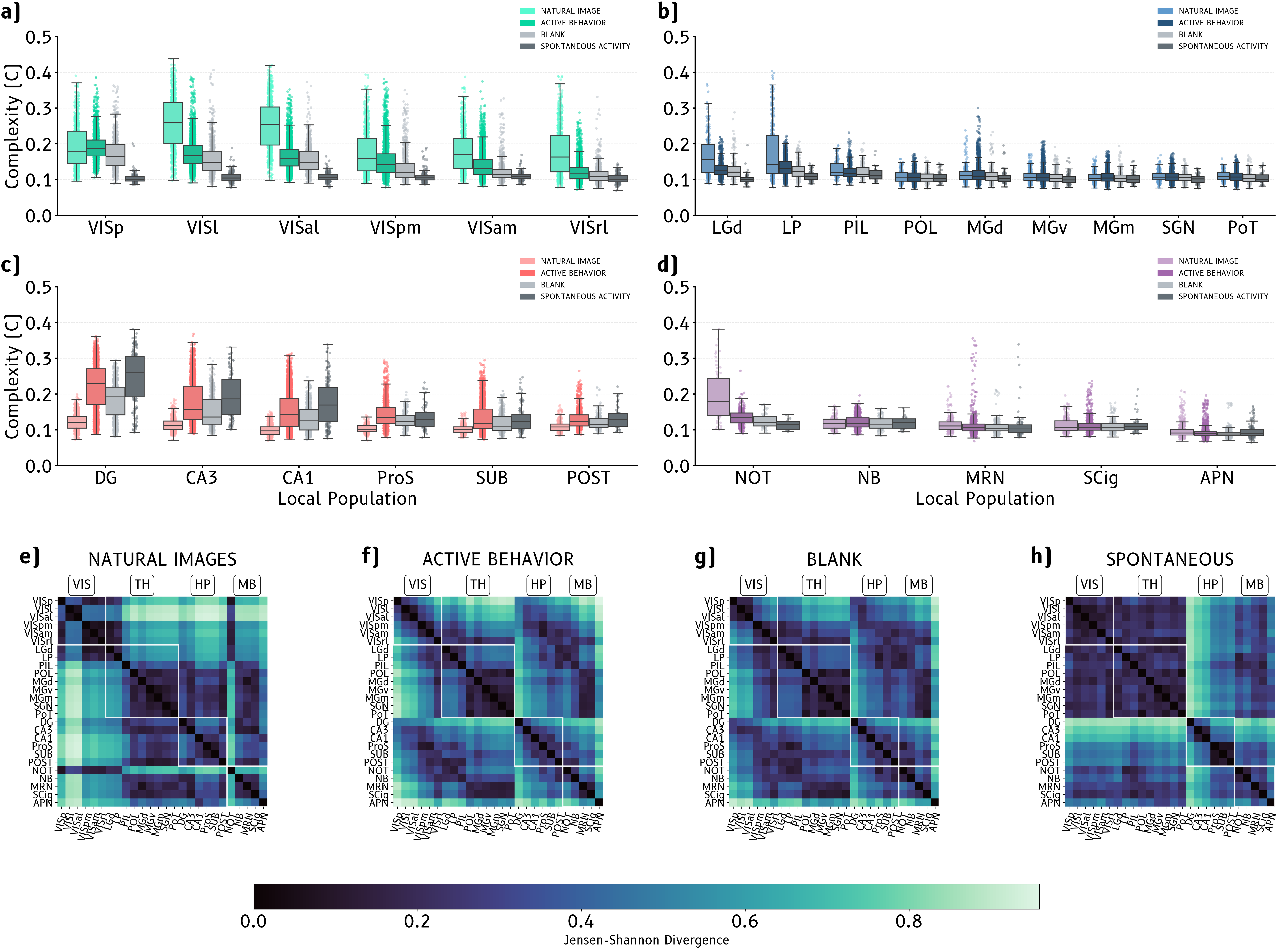}
    \caption{
    \textbf{Statistical complexity exhibits stimulus-dependent modulation in cortical but not subcortical regions.}
    \textbf{(a--d)} Distribution of statistical complexity values across experimental conditions for all brain regions ($n = 102$ sessions, total 77,216 analysis windows). Boxplots show complexity distributions during four behavioral states: natural images (light-colored boxes, $n = 11,321$ windows), active behavior (dark-colored boxes, $n = 51,704$ windows), blank screens (light-gray boxes, $n = 11,321$ windows), and spontaneous activity (dark-gray boxes, $n = 2,870$ windows). 
    \textbf{(a)} Visual cortex subregions exhibited pronounced condition-dependent modulation, with elevated median complexity during visually driven states (VISl: 0.15--0.26; VISal: 0.15--0.25; VISp: 0.16--0.19) compared to spontaneous activity (VISl: 0.1; VISal: 0.1; VISp: 0.1). Natural image presentation showed the highest complexity values, while active behavior and blank screen conditions produced intermediate values. 
    \textbf{(b)} Thalamic regions showed moderate stimulus-dependent modulation with smaller effect sizes (e.g., LP median complexity: 0.12--0.14 during visually driven states vs. 0.11 during spontaneous activity; LGd: 0.12--0.15 vs. 0.1). 
    \textbf{(c)} Hippocampal subregions displayed heterogeneous patterns: DG and CA3 showed low complexity median for natural image presentation (DG: 0.12; CA3: 0.11; CA1: 0.1) and elevated complexity across other conditions (DG: 0.19--0.26; CA3: 0.13--0.18; CA1: 0.12--0.17) with modest condition-dependence, while POST, ProS, and SUB maintained more stable values (0.1--0.13) across conditions. 
    \textbf{(d)} Midbrain areas exhibited the most invariant complexity distributions across conditions (NOT: 0.11--0.18; NB: 0.11--0.12; MRN: 0.10--0.11; SCig: 0.1--0.11; APN: 0.9), confirming stimulus-independent dynamics consistent with CV analysis.
    \textbf{(e--h)} Jensen--Shannon (JS) divergence matrices quantifying dissimilarity between complexity distributions across pairs of brain regions for each experimental condition. Color scale ranges from 0.0 (dark blue, identical distributions) to 1.0 (cyan, maximally divergent distributions). White boxes delineate major anatomical groupings (VIS: visual cortex; TH: thalamus; HP: hippocampus; MB: midbrain). 
    \textbf{(e)} During natural image presentation, visual cortex regions showed high within-group similarity (low JS divergence, $< 0.2$) but large divergences from other brain areas (JS $> 0.4$ for VIS-TH, VIS-MB comparisons), reflecting distinct dynamical regimes. Hippocampal DG and CA3 exhibited moderate divergence from other regions (JS: 0.3--0.5). 
    \textbf{(f)} Active behavior condition revealed similar patterns, with visual cortex maintaining distinct complexity distributions from subcortical structures (JS $> 0.4$). Within-hippocampus divergences increased (JS: 0.2--0.4), reflecting heterogeneity in condition-dependent responses. 
    \textbf{(g)} Blank screen presentation showed comparable divergence patterns to active behavior, with visual cortex remaining divergent from other regions (JS $> 0.4$) despite absence of structured visual input. 
    \textbf{(h)} Spontaneous activity exhibited reduced overall divergences, with visual cortex showing smaller differences from other regions (JS: 0.2--0.4) compared to visually driven states, suggesting convergence toward more similar dynamical regimes in the absence of sensory drive. Midbrain areas showed consistently low divergence across all conditions (JS $< 0.2$ within MB group), confirming their stable dynamics. 
    }
    \label{fig:cplx-stimuli}
\end{figure*}

Having established that statistical complexity captures meaningful aspects of population dynamics that complement CV analysis, we next investigated whether complexity measures also reveal stimulus-dependent modulation across brain regions. If the distinct dynamical regimes observed in Fig.~\ref{fig:cplx-intro} reflect intrinsic network properties independent of sensory context, complexity distributions should remain stable across stimulus conditions. Alternatively, if dynamical complexity is flexibly modulated by behavioral state and visual input, as we observed for CV, regions should exhibit condition-dependent shifts in their positions within the C-H plane. This distinction is particularly important because complexity and CV capture different aspects of population dynamics: while CV reflects the balance between synchrony and asynchrony in population firing, complexity quantifies the temporal structure and predictability of population activity. To test whether stimulus-dependence is a general property of cortical dynamics or specific to particular measures of variability, we compared complexity distributions across the same four experimental conditions examined in Fig.~\ref{fig:cv-stimuli}: natural image presentation, active behavior, blank screens, and spontaneous activity (Fig.~\ref{fig:cplx-stimuli}).

Visual cortex subregions exhibited pronounced stimulus-dependent modulation of statistical complexity [Fig.~\ref{fig:cplx-stimuli}(a)]. All six visual areas (VISl, VISal, VISp, VISam, VISrl, VISpm) showed elevated median complexity during natural image presentation (0.16--0.26) compared to spontaneous activity (0.1), with active behavior and blank screen conditions producing intermediate values (0.1--0.19). This hierarchical organization---highest complexity during structured visual input, intermediate during task engagement, lowest during spontaneous activity---was consistent across all visual areas (all pairwise comparisons $p < 0.001$). Notably, the range of complexity values and their variability (interquartile ranges: 0.08--0.12) was greatest during visually driven states, reflecting heterogeneous temporal structure across different stimuli and behavioral contexts. This systematic modulation demonstrates that visual cortical population dynamics flexibly adjust their temporal organization in response to both sensory input and behavioral state.

Thalamic nuclei displayed more limited stimulus-dependence [Fig.~\ref{fig:cplx-stimuli}(b)]. Among nine examined regions, only the lateral geniculate dorsal (LGd) and lateral posterior (LP) nuclei, both components of the visual thalamic pathway, showed significant modulation across conditions ($p < 0.001$). LGd and LP exhibited elevated complexity during visually driven states (0.14--0.18) compared to spontaneous activity (0.11--0.12), with effect sizes smaller than those observed in visual cortex. The remaining seven thalamic nuclei (MGd, PoT, SGN, POL, MGm, MGv, PIL) maintained uniformly low complexity (0.10--0.13) with minimal variability across all conditions, showing no significant stimulus-dependence. This functional segregation suggests that only primary sensory relay and higher-order visual thalamic nuclei encode stimulus-related temporal structure, while other thalamic regions maintain stable dynamics independent of visual context.

Hippocampal subregions displayed heterogeneous patterns that diverged from the cortical profile [Fig.~\ref{fig:cplx-stimuli}(c)]. DG and CA3 showed elevated complexity across most conditions (DG: 0.19--0.27; CA3: 0.15--0.21) with modest condition-dependence, maintaining relatively high temporal structure regardless of visual input. Intriguingly, these regions exhibited peak complexity during spontaneous activity, opposite to the pattern observed in visual cortex. In contrast, CA1, POST, ProS, and SUB maintained more stable, lower complexity values (0.11--0.15) across states, suggesting distinct computational roles within the hippocampal formation.  

Midbrain areas exhibited the most invariant dynamics across conditions [Fig.~\ref{fig:cplx-stimuli}(d)]. Most regions (NB, MRN, SCig, APN) showed uniformly low complexity (0.10--0.12) with minimal variability across all behavioral states, confirming stimulus-independent dynamics consistent with CV analysis. Only the nucleus of the optic tract (NOT) displayed modest stimulus-dependence (0.13--0.15 during visually driven states vs. 0.13 during spontaneous activity), consistent with its role in visual motion processing. This pattern reinforces the hierarchical organization wherein regions closer to sensory input show flexible, stimulus-dependent complexity, while regions involved in arousal, motor control, and autonomic regulation maintain stable intrinsic dynamics.

To quantify the distinctiveness of dynamical regimes across regions and conditions, we computed Jensen--Shannon divergences between complexity distributions [Fig.~\ref{fig:cplx-stimuli}(e--h)]. During natural image presentation [Fig.~\ref{fig:cplx-stimuli}(e)], visual cortex regions showed high within-group similarity (JS divergence $< 0.2$) but large divergences from other brain areas (JS $> 0.4$ for VIS-TH and VIS-MB comparisons), confirming distinct dynamical regimes. Hippocampal DG and CA3 exhibited moderate divergence from other regions (JS: 0.3--0.5), reflecting their elevated complexity. During active behavior [Fig.~\ref{fig:cplx-stimuli}(f)] and blank screens [Fig.~\ref{fig:cplx-stimuli}(g)], similar patterns persisted, with visual cortex maintaining distinct complexity distributions from subcortical structures. Critically, spontaneous activity [Fig.~\ref{fig:cplx-stimuli}(h)] exhibited reduced overall divergences, with visual cortex showing smaller differences from other regions (JS: 0.2--0.4) compared to visually driven states. This convergence suggests that different brain regions operate in more similar dynamical regimes during spontaneous activity, with visual cortex transitioning from highly structured temporal organization toward simpler dynamics that more closely resemble subcortical baseline states. Midbrain areas showed consistently low divergence across all conditions (JS $< 0.2$ within MB group), confirming their stable dynamics regardless of behavioral context.

These complexity analyses reveal a hierarchical organization of stimulus-dependent dynamics that closely parallels the patterns observed with CV measurements. Visual cortex and primary sensory thalamic nuclei exhibit strong stimulus-dependence, flexibly modulating their temporal structure in response to visual input and behavioral engagement. Hippocampal regions show divergent patterns, with some subfields (DG, CA3, CA1) maintaining high complexity that peaks during spontaneous activity, potentially reflecting internally generated processes. Midbrain regions maintain stable, low complexity dynamics largely independent of sensory context. Importantly, the convergence of complexity distributions during spontaneous activity---with visual cortex approaching subcortical baseline values---suggests that stimulus-driven states actively structure cortical temporal organization, while spontaneous activity reflects a more homogeneous dynamical regime shared across brain regions~\cite{Fox2005}. The concordance between CV and complexity analyses, despite these measures capturing different aspects of population dynamics, indicates that stimulus-dependence is a fundamental organizing principle of cortical function that manifests across multiple temporal scales and statistical properties~\cite{Stitt2019, Buonomano2009}.

\subsection{Complexity-variability coupling reveals state-dependent reconfiguration of cortical dynamics}

\begin{figure}[H]
    \centering
    \includegraphics[width=\columnwidth]{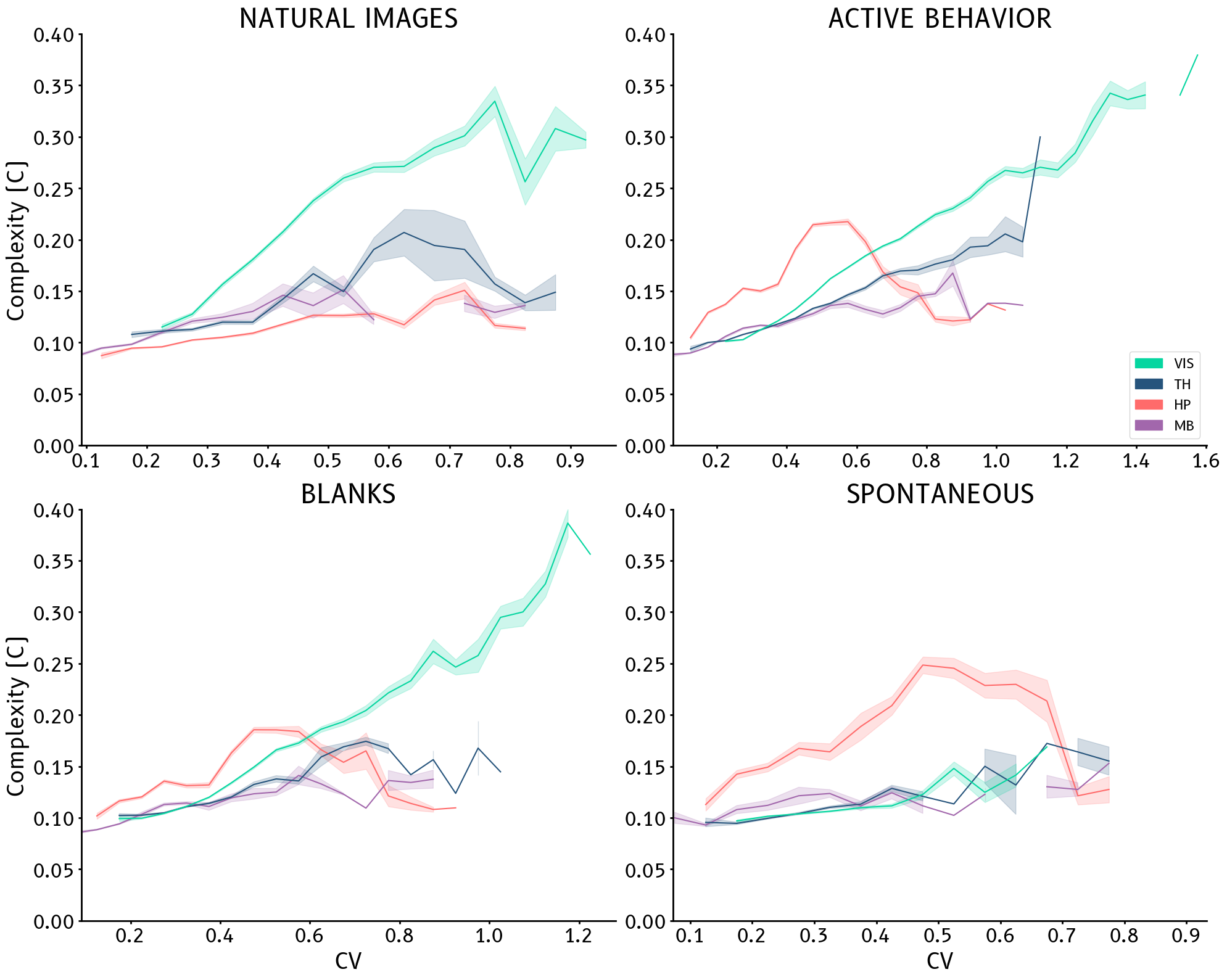}
    \caption{
    \textbf{Relationship between statistical complexity and coefficient of variation across experimental conditions reveals region-specific dynamics.}
    Scatter plots showing statistical complexity ($C$) as a function of coefficient of variation (CV) for four experimental conditions: \textbf{natural images} (top left), \textbf{active behavior} (top right), \textbf{blank screens} (bottom left), and \textbf{spontaneous activity} (bottom right). Lines represent mean trajectories with shaded areas indicating standard error for visual cortex (VIS, green), thalamus (TH, blue), hippocampus (HP, coral), and midbrain (MB, purple) across 102 recording sessions. 
    During natural image presentation, visual cortex exhibited a monotonic increase in complexity with increasing CV (spanning CV: 0.1--0.9, $C$: 0.08--0.34), indicating that higher synchronization (higher CV) corresponds to more structured temporal organization. Thalamus and hippocampus showed moderate increases ($C$: 0.10--0.20), while midbrain displayed minimal modulation ($C$: 0.10--0.15). 
    Active behavior condition revealed similar patterns but with extended dynamic ranges, particularly for visual cortex (CV: 0.2--1.6, $C$: 0.10--0.38), reflecting the broadest repertoire of dynamical states during task engagement. 
    Blank screen presentation showed intermediate patterns (VIS: CV: 0.1--1.2, $C$: 0.10--0.39), with visual cortex maintaining strong complexity-CV coupling despite absence of structured visual input. 
    Spontaneous activity exhibited compressed CV ranges across all regions (CV: 0.1--0.9), with hippocampus showing elevated complexity values ($C$: 0.12--0.25) at moderate CV levels (0.4--0.7), potentially reflecting internally generated dynamics. Visual cortex showed reduced maximal complexity during spontaneous activity ($C \approx 0.15$) compared to visually driven states, while midbrain remained invariant ($C$: 0.10--0.13) across all CV values. These condition-specific trajectories in the $C$-CV plane demonstrate that the relationship between population synchronization and temporal complexity is not fixed, but rather depends on both brain region and behavioral context, with cortical areas showing flexible, state-dependent coupling while subcortical regions maintain more stereotyped dynamics.
    }
    \label{fig:cplx-v-cv}
\end{figure}

Our analyses have demonstrated that both CV and statistical complexity exhibit stimulus-dependent modulation in cortical but not subcortical regions. However, these measures capture fundamentally different aspects of population dynamics: CV quantifies the balance between synchronous and asynchronous firing, while complexity reflects the temporal structure and predictability of activity patterns. A critical question emerges: how do these two dimensions of population variability relate to each other, and does this relationship depend on behavioral state and brain region? If CV and complexity measure independent aspects of neural dynamics, their relationship should remain constant across conditions. Alternatively, if they reflect coupled but dissociable features of network organization, the complexity-CV relationship might reconfigure in a state-dependent manner. To address this question, we examined the joint distribution of complexity and CV values across all experimental conditions (Fig.~\ref{fig:cplx-v-cv}).

Visual cortex exhibited a robust positive relationship between complexity and CV during all visually driven conditions [Fig.~\ref{fig:cplx-v-cv}]. During natural image presentation, complexity increased monotonically with CV, spanning a broad range (CV: 0.1--0.9, $C$: 0.08--0.34). This positive coupling indicates that periods of higher population synchronization (high CV) correspond to more structured, predictable temporal organization (high complexity), while desynchronized states (low CV) exhibit simpler, more irregular dynamics (low complexity). Importantly, visual cortex explored the largest region of the $C$-CV plane, suggesting a rich repertoire of dynamical states. Active behavior further expanded this range (CV: 0.2--1.6, $C$: 0.10--0.38), with visual cortex achieving the highest complexity values at elevated synchronization levels. This extended dynamic range during active behavior may reflect enhanced temporal coordination required for task engagement and rapid stimulus processing. Blank screen presentation showed intermediate patterns (CV: 0.1--1.2, $C$: 0.10--0.39), demonstrating that visual cortex maintains strong complexity-CV coupling even during brief periods without structured visual input, consistent with persistent engagement of cortical circuits during the task.

Thalamic regions displayed weaker but consistent positive complexity-CV relationships across visually driven conditions [Fig.~\ref{fig:cplx-v-cv}]. The slopes of these trajectories were shallower than in visual cortex, and the explored regions of the $C$-CV plane were more restricted (typical range: CV: 0.2--0.8, $C$: 0.10--0.20). This compressed dynamic range suggests that thalamic circuits maintain more constrained relationships between synchronization and temporal structure, consistent with their role as relatively feedforward sensory relays. Nevertheless, the preserved positive coupling indicates that thalamic dynamics are not entirely stereotyped, but retain some flexibility in coordinating temporal patterns with population synchronization levels.

Hippocampal dynamics revealed strikingly different patterns depending on behavioral state [Fig.~\ref{fig:cplx-v-cv}]. During natural images, active behavior, and blank screens, hippocampus showed modest positive complexity-CV relationships with moderate complexity values (typical $C$: 0.12--0.19). However, during spontaneous activity, hippocampus exhibited markedly elevated complexity ($C$: 0.12--0.25) at intermediate CV values (0.4--0.7), reaching levels comparable to those observed in visual cortex during active image viewing. This inversion, with hippocampal complexity peaking during spontaneous rather than stimulus-driven states, suggests fundamentally different computational demands. The elevated complexity during spontaneous activity may reflect internally generated processes such as memory replay, consolidation, or planning that require structured temporal organization independent of external sensory drive~\cite{Fox2005}. This dissociation between hippocampal and cortical dynamics reinforces the view that spontaneous activity serves distinct functions across brain regions, with sensory cortex requiring external input for temporal structure while hippocampus generates rich dynamics intrinsically.

Midbrain regions displayed the most invariant dynamics across all conditions [Fig.~\ref{fig:cplx-v-cv}]. Complexity values remained consistently low ($C$: 0.10--0.13) across the full range of observed CV values (0.1--0.9) in all behavioral states, yielding nearly flat trajectories in the $C$-CV plane. This independence between complexity and CV indicates that midbrain population dynamics maintain simple temporal organization regardless of synchronization level or behavioral context. The restricted exploration of the $C$-CV plane confirms that midbrain circuits operate with stable, stimulus-independent dynamics that likely reflect their roles in arousal, motor control, and autonomic regulation rather than flexible sensory or cognitive processing.

Comparing trajectories across conditions revealed condition-specific reconfiguration of the complexity-CV relationship [Fig.~\ref{fig:cplx-v-cv}]. Visual cortex showed systematic shifts: during visually driven states (natural images, active behavior, blank screens), complexity increased steeply with CV, exploring high-complexity/high-CV regions of the plane. During spontaneous activity, the trajectory compressed, with visual cortex occupying predominantly low-complexity regions ($C < 0.15$) even at moderate CV values. This compression suggests that visual cortex requires sensory input or task engagement to generate highly structured temporal patterns; in the absence of external drive, cortical dynamics simplify despite maintaining some degree of synchronization. In contrast, hippocampus showed the opposite pattern: trajectories expanded during spontaneous activity, with elevated complexity at intermediate CV values. This divergence between cortical and hippocampal state-dependent reconfiguration supports the view that these regions serve complementary functions, with cortex optimized for processing external sensory information and hippocampus specialized for internally generated dynamics~\cite{Buonomano2009}.

The state-dependent reorganization of complexity-CV relationships provides new insights into how brain regions flexibly coordinate temporal structure and population synchronization. The strong, positive coupling in visual cortex during active states suggests that these two features of population dynamics, synchronization and temporal organization, are jointly modulated to optimize sensory processing. The compression of this relationship during spontaneous activity indicates that cortical circuits cannot sustain high temporal complexity without external drive or task demands. The inverted pattern in hippocampus, with peak complexity during spontaneous states, reveals that this region maintains rich temporal dynamics independent of sensory input, potentially supporting offline memory processes. Finally, the flat trajectories in midbrain confirm that these regions maintain stereotyped dynamics with minimal flexibility. Together, these findings demonstrate that the coupling between different dimensions of population variability is not fixed but rather reconfigures in a region-specific and state-dependent manner, providing a mechanistic basis for the functional specialization observed across the visual processing hierarchy.

\section{Discussion} \label{sec:discussion}

\subsection{Hierarchical organization of population dynamics across the visual processing pathway}

Our results reveal a systematic hierarchical organization of population dynamics that extends from cortical to subcortical structures during visual processing. By combining coefficient of variation and information-theoretic complexity measures, we demonstrate that distinct brain regions operate in fundamentally different dynamical regimes, with these differences being most pronounced during sensory-driven states. Critically, the relationship between these two dimensions of population variability---synchronization (CV) and temporal structure (complexity)---reconfigures in a region-specific and state-dependent manner, providing a mechanistic basis for functional specialization across the visual hierarchy.

Visual cortex exhibited the strongest stimulus-dependent modulation across both measures, with CV values increasing 2--3$\times$ during active behavior compared to spontaneous activity, accompanied by systematic shifts in the complexity-CV relationship. During visually driven states, visual cortex explored an extensive region of the C-CV plane (CV: 0.1--1.6, $C$: 0.08--0.38), with complexity increasing monotonically with synchronization. This positive coupling indicates that cortical circuits coordinate both dimensions of population dynamics to optimize sensory processing: elevated synchronization (high CV) coincides with structured temporal organization (high complexity), while desynchronized states exhibit simpler dynamics~\cite{Renart2010, Harris2011}. The compression of this trajectory during spontaneous activity---with visual cortex occupying predominantly low-complexity regions despite moderate CV values---demonstrates that cortical circuits cannot sustain rich temporal structure without external drive or task engagement. This state-dependent reconfiguration aligns with theoretical frameworks proposing that cortical networks flexibly transition between operating regimes based on computational demands~\cite{Gautam2015}.

In contrast, midbrain regions displayed remarkably flat trajectories in the C-CV plane, maintaining low complexity ($C$: 0.10--0.13) across all CV values and behavioral states. This independence between synchronization and temporal structure confirms that midbrain circuits operate with stereotyped dynamics regardless of sensory context, consistent with their roles in arousal and motor control rather than flexible information processing. The hierarchical gradient in both stimulus-dependence and complexity-CV coupling---strongest in visual cortex, intermediate in thalamus, minimal in midbrain---suggests that computational flexibility decreases systematically with distance from primary sensory input.

\subsection{Divergent hippocampal dynamics reveal complementary computational strategies}

Perhaps our most striking finding is the inverted relationship between behavioral state and dynamical complexity in hippocampal circuits. While visual cortex showed compressed C-CV trajectories during spontaneous activity, hippocampus exhibited expanded trajectories with elevated complexity ($C$: 0.12--0.25) at intermediate CV values (0.4--0.7)---comparable to visual cortex during active image viewing. This inversion demonstrates that hippocampal and cortical circuits employ fundamentally different computational strategies: visual cortex requires external sensory drive to generate structured temporal patterns, while hippocampus maintains rich dynamics intrinsically, independent of visual input.

The elevated complexity in hippocampal subregions DG and CA3 during spontaneous activity likely reflects ongoing memory-related processes such as replay, pattern completion, or consolidation that operate autonomously from active sensory processing~\cite{Wilson1994}. The anticorrelated CV relationships between DG/CA3 and visual cortical regions further support this interpretation, suggesting a competitive or complementary relationship between externally-driven sensory processing and internally-generated hippocampal dynamics. This dissociation provides a potential mechanistic explanation for the well-documented suppression of hippocampal replay events during active exploration and sensory engagement.

Importantly, the divergent state-dependent trajectories in the C-CV plane reveal that hippocampus and visual cortex optimize different dimensions of population dynamics under different conditions. Visual cortex achieves high complexity through stimulus-driven synchronization during active behavior, while hippocampus generates high complexity through internally coordinated patterns during quiescence. This complementarity may reflect a fundamental trade-off in neural computation: circuits cannot simultaneously maintain both stimulus-driven responses and internally generated structured dynamics at high levels. The brain appears to solve this by specializing different regions for different computational modes, with sensory cortices prioritizing external sampling and hippocampus prioritizing internal model maintenance~\cite{Fox2005, Buonomano2009}.

\subsection{Complexity-variability coupling as a window into computational flexibility}

The joint analysis of CV and statistical complexity provides novel insights that neither measure alone could reveal. While CV quantifies the balance between synchronous and asynchronous firing, complexity captures the temporal structure and predictability of activity patterns. Our finding that the relationship between these measures is neither fixed nor universal, but rather reconfigures in a region-specific and state-dependent manner, has important implications for understanding cortical computation.

The strong positive coupling in visual cortex during active states (steep slopes in C-CV space) indicates that synchronization and temporal structure are jointly modulated, potentially through common circuit mechanisms such as recurrent excitation or neuromodulatory drive~\cite{McCormick2020}. The compression of this relationship during spontaneous activity suggests that these mechanisms are only engaged by sensory input or task demands. In contrast, the shallow slopes in thalamus and the near-zero slopes in midbrain indicate progressively weaker coupling between these dynamical dimensions, reflecting more constrained circuit architectures with limited flexibility.

The state-dependent reorganization of C-CV trajectories also reveals that population dynamics cannot be adequately characterized by either measure in isolation. Two populations with identical CV values can exhibit vastly different complexity depending on brain region and behavioral state, and conversely, similar complexity values can arise from very different synchronization levels. This highlights a general principle: neural population dynamics are inherently multidimensional, and capturing their full computational capacity requires considering multiple complementary features simultaneously~\cite{Cunningham2014, Gallego2017}.

Moreover, our results demonstrate that the boundaries of dynamical flexibility---the regions of the C-CV plane that each population can explore---are region-specific and behaviorally gated. Visual cortex has the capacity to explore extensive regions but only does so during engaged states. Hippocampus can access high-complexity states but preferentially does so during spontaneous activity. Midbrain remains confined to a restricted region regardless of state. These boundaries likely reflect fundamental constraints imposed by circuit architecture, connectivity patterns, and intrinsic cellular properties~\cite{Mante2013}.

\subsection{Implications for theories of cortical computation and spontaneous activity}

Our findings have important implications for theories of cortical function. The strong stimulus-dependence of cortical dynamics challenges purely intrinsic models of brain function~\cite{Fiser2004} and supports hybrid frameworks where sensory input actively structures cortical dynamics~\cite{Buonomano2009, Stitt2019}. However, the preservation of some temporal structure during spontaneous activity, combined with the elevated hippocampal complexity in this state, suggests that spontaneous activity is not simply noise but rather reflects a distinct computational mode optimized for internal processing~\cite{Fox2005}.

The hierarchical gradient in stimulus-dependence---strongest in primary sensory areas, intermediate in thalamus, minimal in midbrain---is consistent with predictive coding theories where higher levels of the processing hierarchy maintain increasingly stable priors~\cite{Clark2013}. However, our results extend these theories by showing that stimulus-dependence manifests not only in response magnitudes but also in the fundamental relationship between synchronization and temporal structure. Predictive coding models must therefore account for state-dependent reconfiguration of dynamical landscapes, not just modulation of firing rates.

The convergence of complexity distributions across regions during spontaneous activity, despite maintained differences in CV, suggests that the brain defaults to a more homogeneous dynamical regime when not engaged in active processing. This convergence may facilitate global coordination and communication between regions~\cite{Vidaurre2017}, providing an optimal baseline state that balances energy consumption with readiness to respond to incoming stimuli. However, the divergent hippocampal pattern indicates that this convergence is not universal, with some regions maintaining or even enhancing their complexity during rest to support offline computations.

\subsection{Future directions}

Our findings open several avenues for future investigation. First, examining how C-CV trajectories change during learning, adaptation, or attention could reveal how experience and cognitive demands reshape the computational properties of neural circuits. If the boundaries of explored dynamical space expand with learning, this would suggest that circuit flexibility itself is plastic. Conversely, if learning primarily reorganizes where populations reside within a fixed dynamical landscape, this would support models where flexibility is an intrinsic property.

Second, investigating the causal relationship between position in C-CV space and behavioral performance could establish whether the observed dynamical flexibility directly supports cognitive function. Optogenetic or pharmacological manipulations that push populations to different regions of the C-CV plane would test whether high-complexity states are necessary for specific computations or whether they are epiphenomenal correlates of other circuit properties.

Third, extending this analysis to include additional visual structures (superior colliculus, pulvinar, prefrontal cortex) and other sensory modalities would test whether the principles we identified generalize across systems. If similar hierarchical organization and state-dependent reconfiguration appear across modalities, this would suggest universal principles of neural computation. Alternatively, modality-specific patterns might reveal how circuit dynamics are optimized for different types of sensory information.

Fourth, examining how C-CV relationships change across development and aging could provide insights into how dynamical flexibility emerges and potentially degrades. If critical periods are characterized by expanded exploration of dynamical space that later becomes restricted, this would link circuit maturation to computational capacity.

Finally, our results motivate the development of new computational models that explicitly account for the joint dynamics of synchronization and temporal structure. Models should aim to reproduce the observed state-dependent reconfiguration of C-CV relationships and explain the mechanistic basis for region-specific trajectories. Such models would advance our understanding of how circuit architecture, neuromodulation, and sensory input interact to determine population dynamics.

\section{Conclusions}

By combining complementary measures of population variability---the coefficient of variation and statistical complexity---we have revealed a rich hierarchical organization of neural dynamics across the visual processing pathway. The key insight from our joint analysis is that the relationship between synchronization and temporal structure is not fixed but rather reconfigures in a region-specific and state-dependent manner. Visual cortex shows flexible coupling that depends on sensory drive, hippocampus inverts this relationship to prioritize internally generated complexity, and midbrain maintains stereotyped dynamics across states. This multidimensional characterization provides new constraints for theories of cortical computation and demonstrates that understanding neural population dynamics requires considering multiple complementary features simultaneously. The state-dependent trajectories in C-CV space represent a new window into how different brain regions flexibly coordinate multiple dimensions of population activity to support their specialized computational functions.

\section*{Acknowledgments}
\nocite{*}
The following funding agencies support the authors:
CNPq (E.V.P 140199/2023-3; P.V.C. 314094/2023-7; M.C. 310712/2014-9, 301744/2018-1 and 425329/2018-6; F.S.M. 402359/2022-4,  314092/2021-8; N.A.P.V., A.J.F. and L.A.A.A.  444713/2024-7);
FACEPE (P.V.C. APQ 0826-1.05/15; N.A.P.V. ARC-0115-3.13/25);
FAPEAL (F.S.M. APQ2022021000015);
L’ORÉAL-UNESCO-ABC For Women In Science (F.S.M.).
This research is supported by INCT-NeuroComp (CNPq Grant 408389/2024-9) and Finep-Neurobeep (Grant 01.24.0124.00). 

~


\bibliographystyle{apsrev4-2}

%

\end{document}